\RequirePackage[]{lineno}

\documentclass[%
article,
reprint,
twocolumn,
superscriptaddress,
amsmath,
amssymb,
aps,pra,
]{revtex4-2}

\usepackage{graphicx}
\usepackage{dcolumn}
\usepackage{bm}
\usepackage{float}
\usepackage{xcolor}

\begin{document}

\title{Taming the snake instabilities in a polariton superfluid}

\author{Ferdinand Claude}%
 \email{ferdinand.claude@lkb.upmc.fr}
\affiliation{%
Laboratoire Kastler Brossel, Sorbonne Universit\'{e}, CNRS, ENS-Universit\'{e} PSL, Coll\`{e}ge de France, 75005 Paris, France
}%
\author{Sergei V. Koniakhin}%
\affiliation{%
Institut Pascal, PHOTON-N2, Université Clermont Auvergne, CNRS, SIGMA Clermont, F-63000 Clermont-Ferrand, France
}%
\affiliation{Alferov University, 8/3 Khlopina street, Saint Petersburg 194021, Russia}
\author{Anne Ma\^{i}tre}
\affiliation{%
Laboratoire Kastler Brossel, Sorbonne Universit\'{e}, CNRS, ENS-Universit\'{e} PSL, Coll\`{e}ge de France, 75005 Paris, France
}%
\author{Simon Pigeon}%
\affiliation{%
Laboratoire Kastler Brossel, Sorbonne Universit\'{e}, CNRS, ENS-Universit\'{e} PSL, Coll\`{e}ge de France, 75005 Paris, France
}%
\author{Giovanni Lerario}%
\affiliation{%
Laboratoire Kastler Brossel, Sorbonne Universit\'{e}, CNRS, ENS-Universit\'{e} PSL, Coll\`{e}ge de France, 75005 Paris, France
}%
\author{Daniil D. Stupin}%
\affiliation{Alferov University, 8/3 Khlopina street, Saint Petersburg 194021, Russia}
\author{Quentin Glorieux}%
\affiliation{%
Laboratoire Kastler Brossel, Sorbonne Universit\'{e}, CNRS, ENS-Universit\'{e} PSL, Coll\`{e}ge de France, 75005 Paris, France
}%
\affiliation{Institut Universitaire de France (IUF), F-75231 Paris, France}
\author{Elisabeth Giacobino}%
\affiliation{%
Laboratoire Kastler Brossel, Sorbonne Universit\'{e}, CNRS, ENS-Universit\'{e} PSL, Coll\`{e}ge de France, 75005 Paris, France
}%
\author{Dmitry Solnyshkov}%
\affiliation{%
Institut Pascal, PHOTON-N2, Université Clermont Auvergne, CNRS, SIGMA Clermont, F-63000 Clermont-Ferrand, France
}%
\affiliation{Institut Universitaire de France (IUF), F-75231 Paris, France}

\author{Guillaume Malpuech}%
\affiliation{%
Institut Pascal, PHOTON-N2, Université Clermont Auvergne, CNRS, SIGMA Clermont, F-63000 Clermont-Ferrand, France
}%

\author{Alberto Bramati}%

\affiliation{%
Laboratoire Kastler Brossel, Sorbonne Universit\'{e}, CNRS, ENS-Universit\'{e} PSL, Coll\`{e}ge de France, 75005 Paris, France
}%
\affiliation{Institut Universitaire de France (IUF), F-75231 Paris, France}

\date{\today}

\begin{abstract}
The  dark solitons observed in a large variety of nonlinear media are unstable against the modulational (snake) instabilities and can break in vortex streets. This behavior has been investigated in nonlinear optical crystals and ultracold atomic gases. However, a deep characterization of this phenomenon is still missing. In a resonantly pumped 2D polariton superfluid, we use an all-optical imprinting technique together with the  bistability of the polariton system to create dark solitons in confined channels. Due to the snake instabilities, the solitons are unstable and break in arrays of vortex streets whose dynamical evolution is frozen by the pump-induced confining potential, allowing their direct observation in our system. A deep quantitative study shows that the vortex street period is proportional to the quantum fluid healing length, in agreement with the theoretical predictions. Finally, the full control achieved on the soliton patterns is exploited to give a proof of principle of an efficient, ultra-fast, analog, all-optical maze solving machine in this photonic platform.
\end{abstract}

\renewcommand{\floatpagefraction}{.85}%

\maketitle

\section{Introduction}
Dark solitons are fundamental localized excitations characterized by a density dip on a homogeneous background that preserve their shape along the propagation in a nonlinear medium because of the balance between the combined effects of diffraction and non-linearity \cite{kivshar1998dark}. They are intrinsically 1-dimensional objects, becoming unstable to transverse modulations as soon as they are embedded in a 2 or 3-dimensional environments. This behavior due to the so called "snake instabilities"\cite{KIVSHAR2000117} has been theoretically predicted and experimentally observed in optics with self-defocusing nonlinear media \cite{kuznetsov_instability_1995,tikhonenko_observation_1996, mamaev_propagation_1996} and more recently in ultracold bosonic  and fermionic gases \cite{anderson_watching_2001,Dutton2001, cetoli_snake_2013}.
In the latter system, the solitons were observed to break in sound excitations and vortex rings, which are 3D dynamically stable structures. 

More generally, if not stabilized by a supersonic flow \cite{kamchatnov2008stabilization}, snake instabilities induce the dark solitons decay in quantized vortex anti-vortex (VA) pairs leading to the formation of quantum vortex streets, a quantum version of von Karman vortex streets. 
Interestingly, the creation of VA pairs in a subsonic flow interacting with a defect has been reported in  time-resolved pulsed experiments in atomic quantum fluids \cite{Kwon2016}.  However, despite several efforts devoted to the study of this phenomenon, the quantitative study of the snake instability dynamics
remained elusive so far in equilibrium quantum fluids.

Driven-dissipative quantum fluids, namely cavity exciton-polaritons \cite{kavokin2} (polaritons), which are 2D photonic modes interacting via their excitonic parts, emerged in the last two decades as a very flexible photonic platform to study quantitatively topological excitations  \cite{miyahara1985abrikosov,sigurdsson2014information,ma2020realization}.
In these systems, with a resonant excitation, quantized vortices and oblique dark solitons are generated when the flow interacts with a static defect \cite{amo2011polariton,Hivet}. However in this case the solitons are stable even at subsonic flows, due to the intrinsic dissipation of the system \cite{kamchatnov2012oblique} and their breaking in vortex streets is not observed. In incoherently pumped polariton condensates, theoretical proposals were made to generate dark solitons\cite{Smirnov2014,Liew2015}, however the very fast relaxation of the solitonic structures is expected to hinder the observation of vortex street formation.

A scheme which emerged recently is based on resonant pumping and the use of the bistability loop of the non-linear polariton system \cite{Baas2004Opticalbistability, Yulin2008, Zhang2012} where, for a given pumping intensity, the density can be either low or high. When high and low density regions are simultaneously present in the system, stationary phase defects of a new type can exist in low-density regions \cite{Aioi2013,Pigeon2017, koniakhin_stationary_2019,lerario2020vortex,lerario2020parallel,maitre_dark-soliton_2020}, where the phase is not fixed by the resonant excitation, because most particles are not directly injected by the laser but diffuse from higher-density regions which represent potential barriers with sharp interfaces. 

In the resonant configuration, the sharpness of the interfaces is limited by the healing length $\xi$ of the quantum fluid and not by the thermal diffusion of the exciton reservoir, as it is the case for non-resonant excitation. The control of the spatial distribution of the phase and intensity of the pump with spatial light modulators (SLM) allows to realize various confining potentials, such as 1D channels, 0D traps \cite{PhysRevB.89.134501}, or circuits made by the combination of both. 

In this work we use an all optical imprinting technique together with the polariton bistability to create dark solitons confined inside 1D elongated channels. We observe the soliton snake instability leading to the formation of symmetric arrays of vortex streets, which are frozen by the pump-induced confining potential allowing their direct observation and a quantitative study of the onset of the instabilities.

Moreover, by exploiting the full optical control and the flexibility offered by our photonic platform we demonstrate its applied potential by implementing an all optical, programmable maze solving machine. 

A maze of arbitrary shape is optically imprinted as a complex set of 1D channels with dead ends and only an entrance and an exit. We observe that in a certain range of parameters the vortex streets disappear from the dead ends and remain only along the path connecting the entrance and exit, then "solving" the maze within a picosecond timescale. This ultra fast all-optical maze solving machine represents an important step for the large interdisciplinary field of analog graph solving algorithms \cite{Shannon1951,Steinbock1995,nair2015maze,Caruso2016,Berloff2017,Chao2019,Adamatzky2020}, opening a new field of applications for the quantum fluids of light.

\section{Experimental setup}
\begin{figure}[!htbp]
\includegraphics[width=0.48\textwidth]{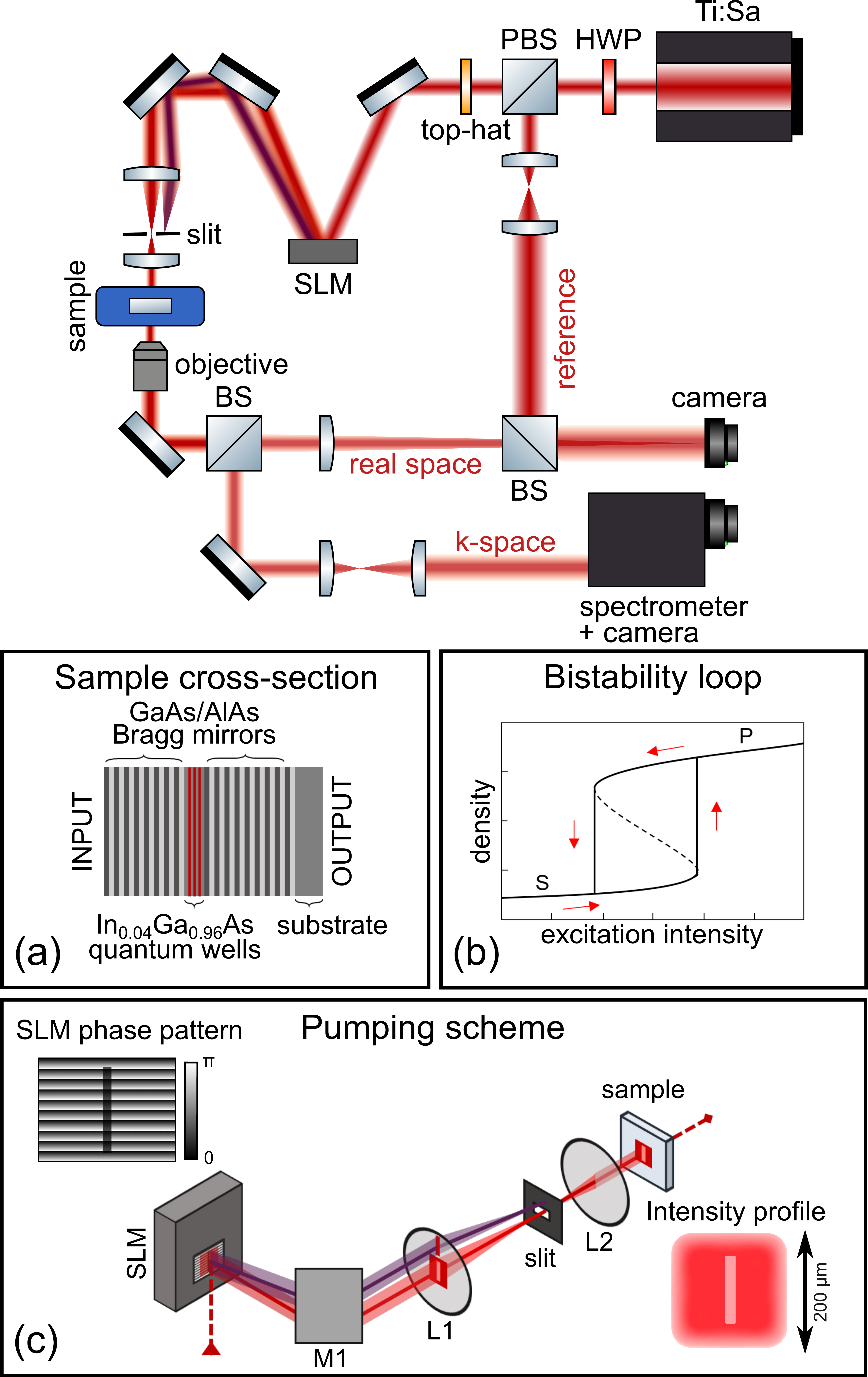}\caption{\label{fig:fig1} Experimental setup. Panel (a) shows a cross-section skecth of the microcavity. Panel (b) is a sketch of the bistability loop obtained with a quasi-resonant excitation. Panel (c) gives in detail the shaping method to dig a rectangular vertical channel in the center of the driving field. In the scheme of the experimental set-up the red beam after the SLM corresponds to the laser intensity diffracted by the grating; the non-diffracted purple beam corresponds to the zero order of the grating, which is cut by a slit in the Fourier plane to obtain the intensity profile shown in the input plane of the cavity. The laser spot is flat and square due to the use of a top-hat lens. The real space detection arm gives access to the density and phase maps of the fluid, while the k-space one to the energy-momentum distribution of the fluid by using a spectrometer.}
\end{figure}

Our sample is a planar 2$\lambda$ GaAs high-finesse semiconductor microcavity made of 21 top and 24 bottom GaAs/AlAs Bragg mirror layer pairs, with one embedded $In_{0.04}Ga_{0.96}$As quantum well (QW) at each of the three antinodes of the confined electromagnetic field (figure \ref{fig:fig1}(a)). QW excitons strongly couple to the first confined radiative mode of the microcavity to form polariton quasi-particles. Polaritons are massive (mass $m$) bosons with repulsive interactions (interaction constant $g$).

The Rabi splitting and the polariton lifetime were measured to be respectively around $\hbar\Omega=5.1$~meV and $\tau_{pol}=15$~ps in experiments conducted in an open-loop flow helium cryostat at a temperature of 5~K. The energy detuning between the photon and exciton modes can be tuned around 0~meV by changing the working point on the microcavity, due to the presence of a small wedge between the cavity mirrors.

A cw single mode Ti-Sapphire laser injects photons at normal incidence with respect to the microcavity plane, so that the fluid has no imposed flow velocity in the transverse plane. It has a small positive energy above the resonant states of the lower polariton branch, typically blue-detuned by $\hbar\omega=0.1$~meV. For homogeneous photonic systems with repulsive (defocusing) interactions, this type of pumping is associated with a hysteresis bistability cycle, as shown in figure \ref{fig:fig1}(b). Starting from an empty system, the rise of the pumping intensity takes place with a weak absorption up to the turning point, where the system jumps on the higher bistability branch. The interaction energy between the created polaritons is there slightly larger than the laser detuning, which gives a good estimate of the height of the induced confining potential. It is then possible to reduce the pumping while staying on the upper branch with a large density of polaritons in the system. 

The confinement potential is obtained by shaping the intensity of the driving field using a spatial light modulator (SLM). The latter displays a blazed diffraction grating, which deviates a part of the laser intensity in a given off-axis direction with respect to the main optical path. The diffraction efficiency depends on the contrast of the grating pattern. Thus, by locally changing the phase modulation of the grating, it is possible to "dig" shapes of low intensity together with phase jumps inside the profile of the laser beam. We call $P$ the pumping density in the high density regions where the system lies on the upper bistability branch and $S$ the pumping density in the low density regions, where the system lies on the lower bistability branch. The spot is previously flattened using a square top-hat diffractive optical lens to get a uniform intensity within the channel.

In the elongated channel case, the surrounding high-density walls are generated by the high phase-contrast regions of the grating, while the low-density channel comes from the low phase-contrast rectangular shape at the center of the SLM (figure \ref{fig:fig1}(c)). Then, by aligning the diffracted order on the optical axis and by cutting the non-diffracted zero order using a slit in the Fourier space, we obtain in the plane of the cavity the desired channel geometry. This all-optical approach turns out to be very flexible for simultaneously designing in the fluid arbitrary confinement potentials and controlling the relative intensity and phase between the external and internal regions of these shapes.

A CCD camera collects the light transmitted by the cavity through the Bragg bottom mirror. The phase measurements are made by interfering the emitted light with a reference beam extracted from the driving field before any modulation by the SLM.

The simulations of the polariton quantum fluid under resonant pumping are carried out using two coupled equations: the Schrodinger equation for the wave function of photons $\psi$ and the Gross-Pitaevskii equation for the wave function of excitons $\chi$. The low wave vector limit of this system of equations corresponds to the Lugiato-Lefever equation \cite{Lugiato1987} well known in nonlinear optics. We neglect the polarization degree of freedom and any thermal effects. The equation reads:
\begin{eqnarray}
i\hbar\frac{\partial \psi }{\partial t}  &=& \left[-\frac{\hbar^2\nabla
^2}{2m} -i \Gamma_P \right]\psi +\hbar\Omega\chi + Pe^{-i \omega_0 t}\\
i\hbar\frac{\partial \chi }{\partial t}  &=& \left[ -i \Gamma_X + g\left| \chi\right|^2 \right]\chi+\hbar\Omega\psi
\label{eq_GPEph}
\end{eqnarray}
where $\Gamma_P =\hbar/(2 \tau_P)$ is the photon decay rate ($\tau_P=7.5$~ps), $\Gamma_X =\hbar/(2 \tau_X)$ is the exciton decay rate ($\tau_X=150$~ps),  $m=4\times 10^{-5}m_0$ is the photon mass ($m_0$ is the free electron mass), $g=7~\mu$eV$\mu$m$^2$ is the exciton-exciton  interaction constant. 

\section{Results and discussion}

\begin{figure}[htbp]
\includegraphics[width=0.47\textwidth]{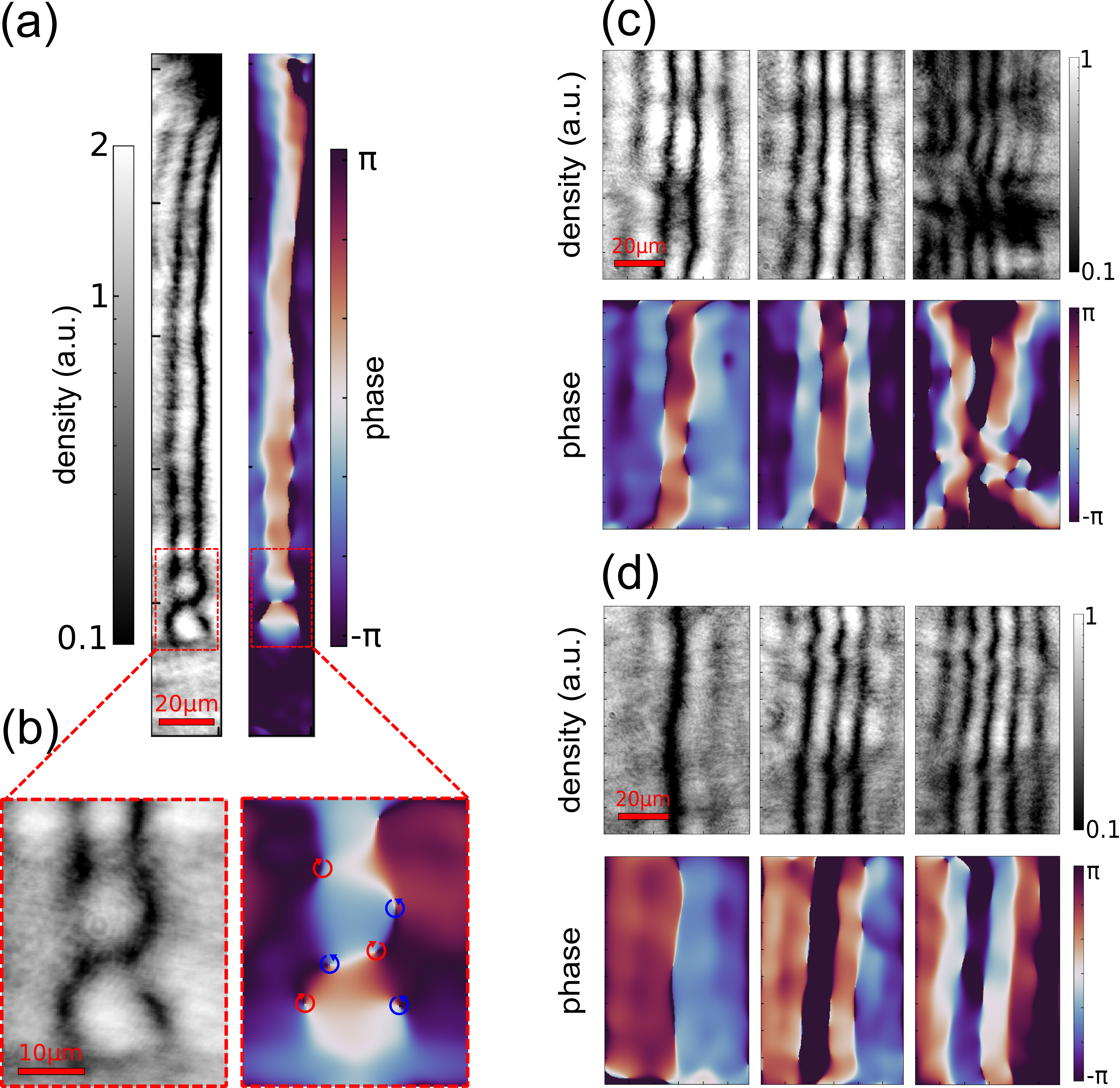}
\caption{\label{fig:fig2} Density and phase of the polariton fluid for different channel geometries. Panel (a) shows a channel of length l = 240 $\mu$m and width L = 15 $\mu$m for a laser detuning $\Delta$E = 0.2 meV. 
A pair of dark solitons spreads along the channel, as confirmed by the two phase jumps of $\pi$. Its decay in VA pairs is observed in the bottom part of the channel, closed by a high-density wall. Panel (b) is a zoom of this region where several winding of 2$\pi$ on the phase map are visible. The panel (c) (panel (d)) shows three open-end channels surrounded by two in phase (out of phase) walls, imprinted with different widths L = 25, 40, 55 $\mu$m for a detuning $\Delta$E = 0.05 meV. The channel contains 2, 4 and 6 (1, 3 and 5) dark solitons.
}
\end{figure}

We start with the study of different configurations of dark solitons in elongated 1D channels (Fig.~\ref{fig:fig2}). The phase patterns, processed by Fourier filtering and phase unwrapping, allow a clear observation of the characteristic $\pi$ phase jump of dark solitons. We have  first investigated the impact of the boundary conditions at the ends of the channel. In figure \ref{fig:fig2}(a), the top end of the channel is open, being linked to the low-density region at the edge of the driving field.  The bottom end is closed by a high-density horizontal wall. The breakdown of the translational symmetry due to the presence of the wall triggers the snake instability and leads to  the formation of VA pairs. Fig.~\ref{fig:fig2}(b) is a zoom on the intensity and phase defects existing near this edge, which clearly shows the 2$\pi$ phase winding of quantum vortices, indicated by red and blue circles. 

In Figure \ref{fig:fig2}(c) we study the case of channels with two open ends and show that by adjusting the width of the channel and the relative phase between the two channel walls, one can precisely tune the number of solitons. Here, for a zero relative phase and for widths of 25~$\mu$m, 40~$\mu$m and 55~$\mu$m, we observe two, four and six dark solitons respectively. For wide channels, as in the case of six solitons, the confinement induced by the walls is no longer sufficient to sustain stable solitons and the modulational instability starts to develop.

The parity of the soliton number is imposed by the relative phase between the channel walls. Indeed, the dark solitons are characterized by a $\pi$ phase shift. In order to ensure the continuity of the quantum fluid wave function, they can appear only by pairs in a channel with zero relative phase between the two walls, which is equivalent to a multiple of 2$\pi$. When the relative phase between the pumps is $\pi$, the soliton number is odd, as shown in Fig.~\ref{fig:fig2}(d). This type of observation was made previously \cite{Goblot2016} in a 1D system. In this case $x$ direction corresponded to the confinement direction, whereas the quasi-free $y$-direction is absent, which makes impossible the development of the modulational instability.

\begin{figure}[htbp]
\centering
  \includegraphics[width = 1\linewidth]{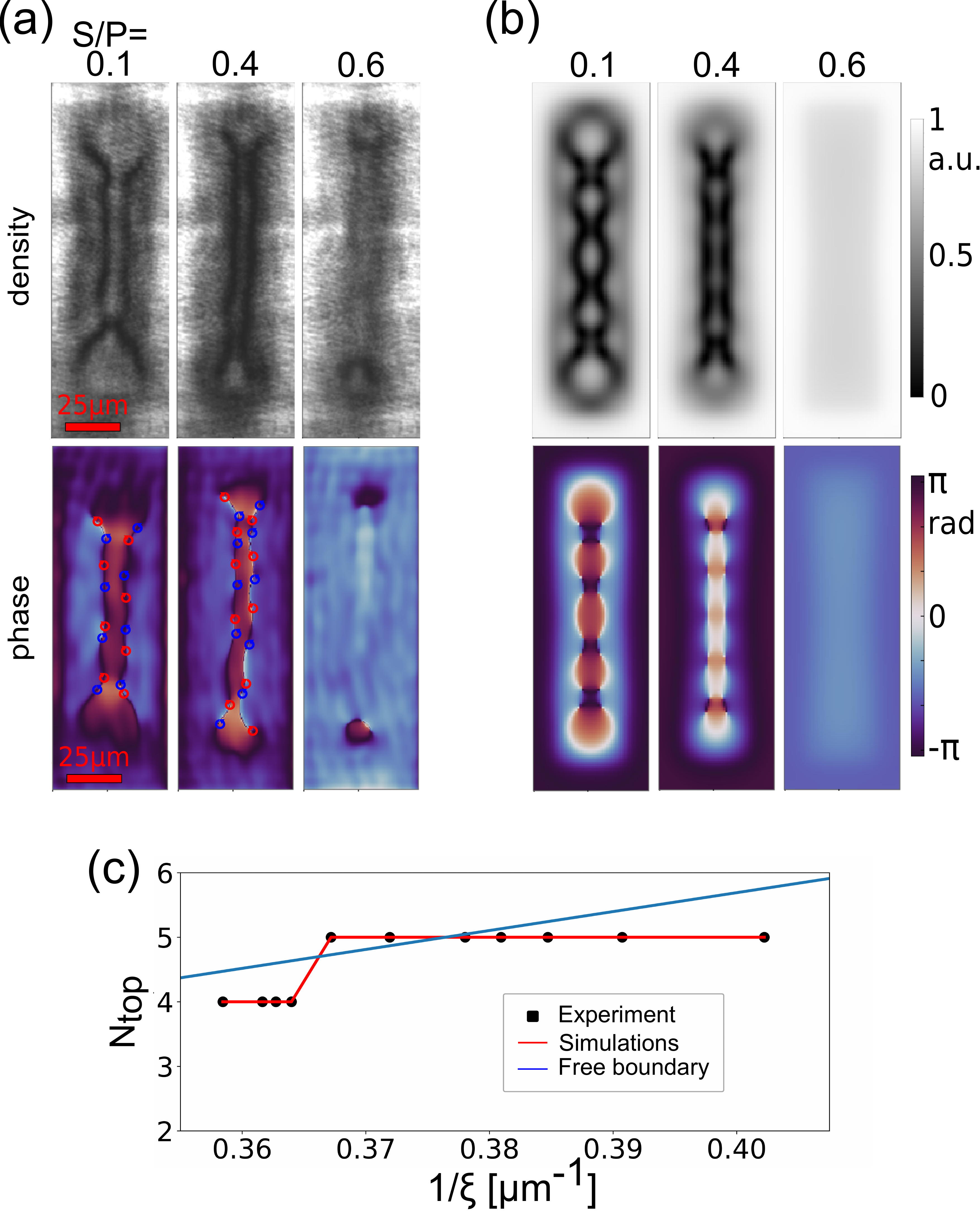}
\caption{\label{fig:fig3} Panel (a): experimental fluid density and phase maps for a channel with length and width  respectively l = 150 $\mu$m, L = 23 $\mu$m; the ratio S/P increases from left to right.
Panel (b): corresponding numerical simulations. Panel (c): Evolution of the number of VA pairs N$_{top}$ in the channel for different values of $1/\xi$. The black dots are the experimental data. The red curve gives the simulations results under the same experimental conditions whereas the blue curve shows the vortex density for an infinite channel.}
\end{figure}

We now consider a channel with both the ends closed by high-density walls (Figure \ref{fig:fig3}). The channel width is 23 $\mu$m, which for our set of parameters allows to generate a single dark soliton pair, which evolves toward a stationary frozen vortex street due to the snake instability.

In figure~\ref{fig:fig3}, the panel (a) shows the measured intensity and phase distributions in the channel for increasing ratios $S/P$.  The panels (b) shows the corresponding numerical simulations obtained by solving the system of equations~\eqref{eq_GPEph}. A symmetric array of VA pairs, similar to a Von-Karmann vortex street is visible for $S/P = 0.1$ and $0.4$, respectively. The slight deviation with respect to the symmetric configuration in the experiment is most probably due to the presence of structural disorder in the microcavity. A soliton pair is indeed unstable in this regime \cite{koniakhin_stationary_2019} against modulational "snake" instability. The soliton pair therefore breaks into the observed VA chain. In free space, these chains would dynamically evolve \cite{Dutton2001} to eventually disappear. While we cannot resolve here the time evolution experimentally (a movie showing an example of this evolution in theory is shown in Visualization 1), remarkably the presence of the confining potential allows to get access to the snake structure, frozen at a given stage of its evolution.

The particle density in the channel is associated with the healing length of the fluid $\xi=\hbar/\sqrt{gn m}$ ($n$ is the density) which sets the dark soliton width and the vortex core size. 

It also naturally sets the spatial period at which the instability develops along the main channel axis \cite{koniakhin_stationary_2019} and therefore the number of VA pairs which appear in a channel of a finite length $L$. The panel (c) of figure~\ref{fig:fig3} shows the number of pairs experimentally observed versus $1/\xi$ and in red, the theoretical value obtained by numerically solving the system of equations~\eqref{eq_GPEph}, which are in excellent agreement. They both show a step-like increase due to the quantization imposed by the finite channel length. The blue line shows the expected number of vortices per length $L$ in an infinite channel which is proportional to $\sim 1/\xi$.

\begin{figure}[htbp]
\centering
  \includegraphics[width=0.90\linewidth]{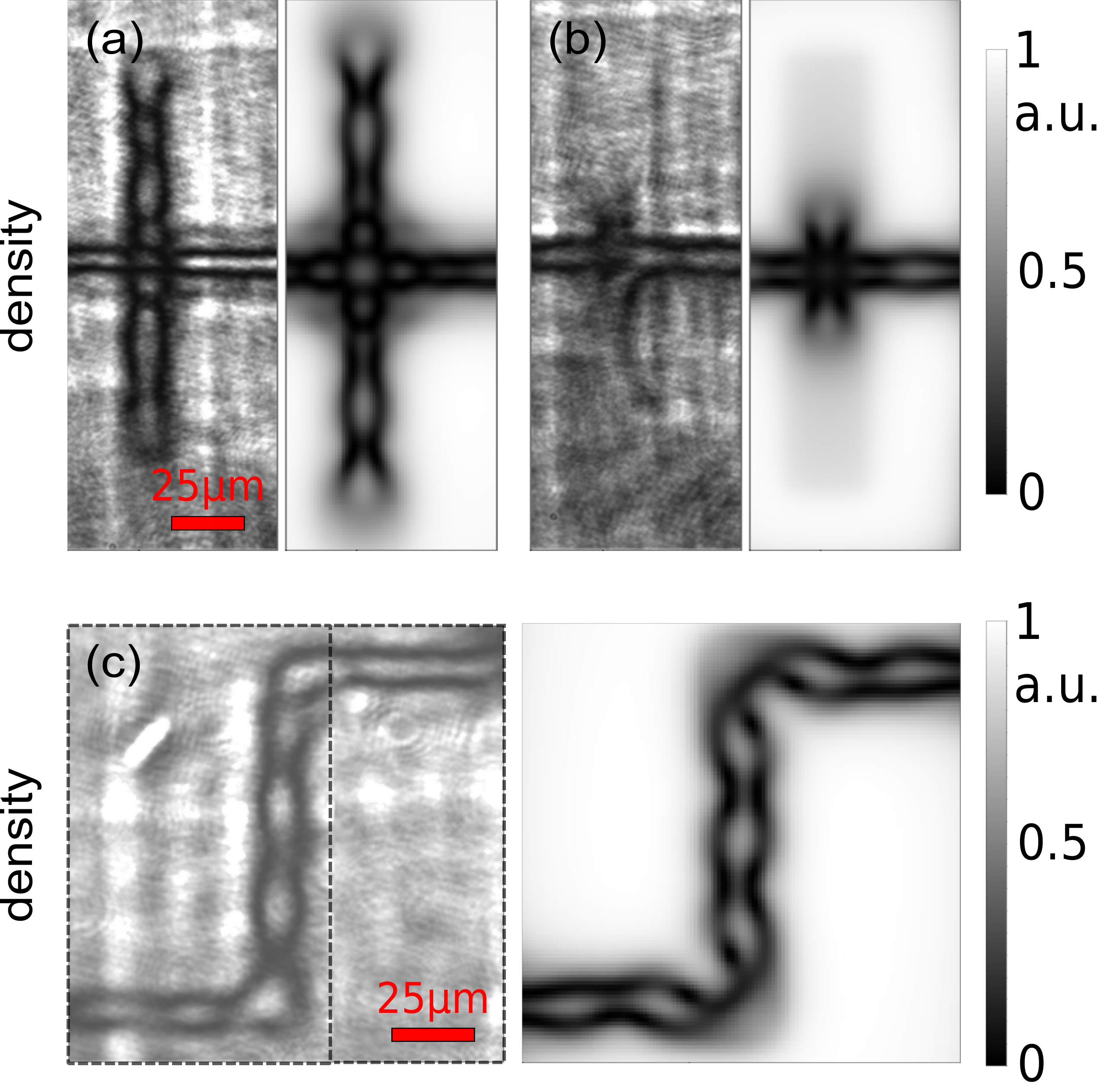}
  \caption{\label{fig:fig5} Panels (a, b): Experimental (left) and numerical (right) density maps of the imprinted cross; S/P = 0.3, channel width L = 23 $\mu$m, laser detuning $\Delta$E = 0.04 meV. In Panel (a) the cavity plane is slightly out of focus: solitonic structures are present in both the cross arms. In panel (b), the cavity plane is in focus, the soliton pair is formed only in the horizontal arm, connecting the entrance with the exit, thus solving the maze. Panel (c): Experimental (left) and numerical (right) density maps of the imprinted S-shaped channel; width $L = 23$~$\mu$m, laser detuning $\Delta E = 0.06$~meV. A dark soliton pair spreads along the whole channel, proving that complex shapes can be used to manipulate and guide the solitons.}
\end{figure}

In Fig~\ref{fig:fig3}, panels (a, b) ($S=0.6P$)  we observe that solitons are no more visible in the channel. 
In this regime, the density inside the channel has increased, therefore the walls are no longer fixed: the channel closes, thereby causing the vanishing of the confinement potential and therefore of the soliton pair.

The critical support value S$_c$  at which the channel gets filled by a homogeneous polariton fluid depends on its geometry. Indeed the transverse domain wall of closed channels is confined by the main walls. The particles belonging to the confined wall possess a larger kinetic energy than the one of an infinite wall, which modifies the pumping densities of the bistability cycle reducing the critical support value at which the wall propagates. A closed channel typically shows a critical value S$_{cc}<S_c$ lower than that of the same open channel at studied in details in \cite{koniakhin_stationary_2019}

This difference can be exploited to solve an analog maze, built with an arrangement of several open and closed channels. First, the maze has an entry and an exit situated in the low-density regions and connected to each other by a channel. Second, it has a set of dead-end paths, modeled by channels closed with a third wall. In this configuration, by judiciously adjusting the $S/P$ ratio and the width of the channels, it is possible to fill only the dead-end paths ($S>S_{cc}$) in such a way that a pair of solitons only passes through the open path (because $S<S_c$) connecting  the entry to the exit and therefore solving the maze.

In figure~\ref{fig:fig5}, panels (a, b) illustrate, both experimentally and theoretically, the proof of principle of the maze resolution. We imprint in the fluid two orthogonal channels crossing each other at the center. The vertical channel is closed while the horizontal one is open. They are both excited at the same intensity $S$, set at $0.3P$. This cross shape mimics a simple maze, in which one searches the path between the entry and the exit represented here by the two open ends of the horizontal channel.

The excitation conditions between the two images are tuned by changing the position of the lens focusing the driving field on the sample. In figure~\ref{fig:fig5}, panel (a), the positions of the input plane of the sample and the laser waist are slightly offset, in figure~\ref{fig:fig5} panel (b), they are coincident. In such a way, we tune the $S_c$ values of the open and closed channels while keeping a constant $S/P$ value.

In the first case, the two channels are both filled with a soliton pair. Then, by bringing the sample closer to the position of the laser waist, which leads simultaneously to an increase of the intensity in the walls/cross regions and a decrease of the width of the channels, we reach an excitation regime without solitons in the closed channel. In the open channel $S_c$ remains higher than $S$ while in the closed channel, $S_{cc}$ is now lower than $S$. Consequently, the walls of the vertical (closed) channel are no longer fixed and the confinement potential vanishes, causing the disappearance of the vertical solitons. In the end, only the open path - here the horizontal channel - contains a soliton pair: the maze is solved.

All-optical maze solving belongs to the  interdisciplinary field of analog graph and maze solving algorithms \cite{Shannon1951,Steinbock1995,Caruso2016,Berloff2017,Tweedy2020}. The first scientific works on the mazes by L. Euler were the foundation of topology.
Maze solving with robots started with the seminal work of Shannon \cite{Shannon1951} and has developed so much that now liquid metal droplets are solving mazes \cite{Adamatzky2020}, paving the way towards future liquid metal robots. The algorithms used in robot maze solving implement the so-called potential method \cite{Pavlov1984} (the destination is assigned a potential, and the choices are determined by its gradient). The same method is naturally used in biological maze solving \cite{Nakagaki2000,Tweedy2020}, where the motion of biological organisms is controlled by gradients as well, although superior animals also use more advanced techniques, such as path integration \cite{Schatz1999}. In chemical methods, the velocity map of the reaction front allows to find the shortest paths between any two points of a maze \cite{Steinbock1995}. In microfluidics, the maze solving uses the pressure gradient between the input and the output \cite{Fuerstman2003} or the voltage gradient in plasma \cite{Reyes2002}. In optics, escaping from the maze can be achieved via the diffusion of a quantum wavepacket \cite{Caruso2016} (here, the maze is not solved, the particle simply spreads over all sites, including the output site).

The maze solving occurring in the present work implements the dead-end filling algorithm, and not the potential method discussed above. However, the filling of the dead ends is still based on the gradients, as discussed in \cite{koniakhin_stationary_2019}.
The solving time is determined by the velocity $v$ of the transverse domain wall, arising from these gradients,  and by the dead end channel length which is $N L$ in the worst case, where $N$ is the number of cells in the maze given by $N=Z^2/L^2$, with $Z$ the overall system size and $L$ the width of a channel. The solving time is therefore $t=NL/v$. The main advantage of this analog solver is the small value of the prefactor $L/v\sim 0.5$~ns: the high velocity $v$ reduces the solving time, allowing to outperform a modern processor which needs hundreds of clock ticks to check a single cell.

Of course, the cross shape we are presenting is a rudimentary architecture of maze. However, we verified that it is also possible to guide solitons through more complex paths, as shown in Fig.~\ref{fig:fig5}, panels (c). Here, we imprint in the fluid an S-shaped channel with  two open ends in which we observe a soliton pair and its decay into vortices, in excellent agreement with the numerical simulations. Thereby, we are not restricted to work with straight channels. These results prove that we have the basic elements for building more complex maze designs. 

In this work, we were mainly limited by the presence of a large number of structural defects in the microcavity, which  modify  locally the potential felt by the polaritons, constraining the positions where we could create identical channels. Also, the lack of laser power necessary to work on a wider imprinting surface, while maintaining both a homogeneous intensity in the different channels and a large enough energy detuning, prevented the realization of larger mazes. Both these parameters can be improved by using better quality samples which are now available \cite{Ballarini2017} and more powerful excitation lasers. After these technical limitations will be overcome, the high flexibility of the all-optical maze imprinting method could open the way to the implementation of a universal maze solving machine \cite{koniakhin_stationary_2019}.

\section{Conclusions}
We have experimentally demonstrated that the modulational instabilities can be controlled and stabilized in a driven-dissipative polariton system allowing the on-demand formation of soliton pairs and vortex streets. This control allowed to quantitatively measure the spatial period at which the instability develops versus the particle density. 
Non-stationary regimes can be used for fast analog maze solving.

\medskip

\noindent\textbf{Funding.} This work has received funding from the French ANR grants ("C-FLigHT" 138678 and "Quantum Fluids of Light",  ANR-16-CE30-0021), from the ANR program "Investissements d'Avenir" through the IDEX-ISITE initiative 16-IDEX-0001 (CAP 20-25), from the European Union Horizon 2020 research and innovation programme under grant agreement No 820392 (PhoQuS).

\medskip

\noindent\textbf{Acknowledgments.} QG, AB, and DS thank the Institut Universitaire de France (IUF) for support. SVK and DDS acknowledge the support from the Ministry of Education and Science of the Russian Federation (0791-2020-0006).

\medskip

\noindent\textbf{Disclosures.} The authors declare no conflicts of interest.

\medskip

\bibliographystyle{apsrev4-2}
\bibliography{main.bib}

\end{document}